\documentclass[aps,twocolumn]{revtex4}

\usepackage{graphicx}
\usepackage{epstopdf}

\begin{document}
\title{Substructural cooperativity and parallel versus sequential events \\during protein unfolding} 

\author{Lothar Reich and Thomas R.\ Weikl\\[0.1cm] 
\small Max Planck Institute of Colloids and Interfaces, Theory Department,
14424 Potsdam, Germany}

\begin{abstract}
According to the `old view', proteins fold along well-defined sequential pathways, whereas the `new view' sees protein folding as a highly parallel stochastic process on funnel-shaped energy landscapes. We have analyzed parallel and sequential processes on a large number of Molecular Dynamics unfolding trajectories of the protein CI2 at high temperatures. Using rigorous statistical measures, we quantify the degree of sequentiality on two structural levels. The unfolding process is highly parallel on the microstructural level of individual contacts. On a coarser, macrostructural level of contact clusters, characteristic parallel and sequential events emerge. These characteristic events can be understood from  loop-closure dependencies between the contact  clusters. A correlation analysis of the unfolding times of the contacts reveals a high degree of substructural cooperativity within the contact clusters.
\end{abstract}
\maketitle

\section*{Introduction}

How do proteins fold into their native structure? In 1968, Levinthal suggested that proteins are guided along sequential pathways into the native structure, since an unguided search of the vast conformational space seemed incompatible with fast and efficient folding \cite{Levinthal68}. About a decade ago, a `new view' \cite{Baldwin94} emerged in which folding is seen as a parallel process on funnel-shaped landscapes, inspired by simple statistical-mechanical models  (for reviews, see \cite{Dill97,Bryngelson95}). The bias of the funnel landscapes towards the native protein structure ensures efficient folding along a multitude of routes. An intriguing question is whether the apparently contradictory `old view' of sequential folding and `new view' of parallel folding can be reconciled \cite{Dill97,Lazaridis97,Pande98}.

We explore here parallel and sequential events during unfolding of CI2. The protein CI2 is a central model system for folding, because of its prominent role as first protein for which two-state kinetics has been observed \cite{Jackson91} and an extensive mutational analysis of the kinetics \cite{Otzen94,Itzhaki95}. The folding kinetics of CI2 has been investigated theoretically both with atomistic \cite{Li94,Li96,Kazimirski01,Day02,Day05,Lazaridis97,Ferrara00a,Ferrara00b,Li01} and simplified statistical-mechanical models \cite{Munoz99,Alm99,Galzitskaya99,Clementi00,Hoang00,Bruscolini02,Ozkan04}. Since atomistic folding simulations are still limited to small or ultrafast folding proteins \cite{Duan98,Ferrara00,Zagrovic02,Snow02,Snow04,Settanni05,Cavalli05}, MD unfolding simulations at elevated temperatures are an important tool to study the kinetics \cite{Li94,Li96,Kazimirski01,Day02,Day05,Lazaridis97,Tirado97,Wang99,Tsai99,Gsponer01,Paci02,Sham02,Ma03,Morra03,Merlino04}. Daggett and coworkers have used MD unfolding simulations of CI2 to characterize the transition-state ensemble \cite{Li94,Li96,Kazimirski01,Day02,Day05,Day05b}. Lazaridis and Karplus \cite{Lazaridis97} and Ferrara et al.\ \cite{Ferrara00a,Ferrara00b} have extracted  characteristic sequences of events from average unfolding times of contacts in MD simulations.
 
\begin{figure}[b]
\resizebox{0.6\linewidth}{!}{\includegraphics{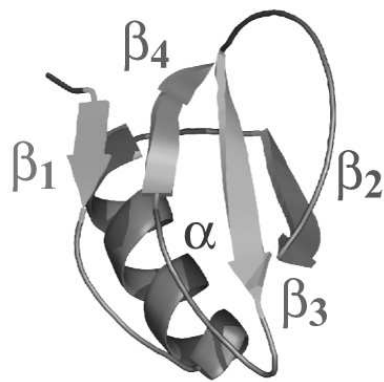}}
\caption{Structure of the protein CI2 \cite{McPhalen87}.}
\label{figure_structure}
\end{figure}

\begin{figure}
\begin{center}
\resizebox{0.7\linewidth}{!}{\includegraphics{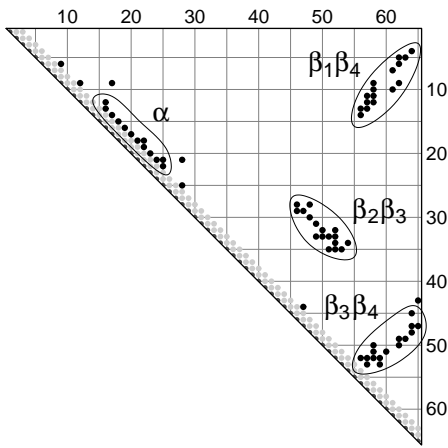}}
\end{center}
\caption{Native contacts and contact clusters of CI2. Black dots represent contacts between pairs of amino acids that where present for at least  75\% of the simulation time on an exemplary MD-trajectory at 300 K starting from the crystal structure. Two amino acids here are defined to be in contact if the distance between their C$_\alpha$ or C$_\beta$ atoms is less than 6\AA.}
\label{figure_contactmap}
\end{figure}

In this article, we quantify the degree of sequentiality during unfolding for each pair of native contacts, and each pair of contact clusters. The contact clusters in the native contact map of CI2 correspond to the four main substructural elements: the $\alpha$-helix, and the three strand pairings $\beta_1\beta_4$, $\beta_2\beta_3$, and $\beta_3\beta_4$ (see Figs.~\ref{figure_structure} and \ref{figure_contactmap}). We consider the sequences of unfolding events of contacts and contact clusters on each trajectory. The unfolding of a pair of contacts or contact cluster is defined as sequential if the same sequence of events is observed on essentially all trajectories. The pairwise unfolding is parallel if contact (or contact cluster) A unfolds prior to contact (or contact cluster) B on some of the trajectories, and later than B on other trajectories.

We find characteristic parallel and sequential unfolding events of the contact clusters. The three contact clusters $\alpha$,  $\beta_2\beta_3$ and $\beta_3\beta_4$ unfold essentially parallel to each other, but sequentially after the contact cluster $\beta_1\beta_4$. This unfolding scenario is in agreement with a simple folding model based on the loop-closure dependencies between the contact clusters \cite{Weikl03a,Weikl05}. In the model, the entropic loop-closure cost for forming the highly nonlocal contact cluster $\beta_1\beta_4$ is reduced by the previous formation of the other three contact clusters. This leads to the prediction that $\beta_1\beta_4$ folds after the other three clusters, i.e.~in reverse sequence to our MD unfolding scenario. In the simple model, the three contact clusters  $\alpha$,  $\beta_2\beta_3$ and $\beta_3\beta_4$ fold in parallel since the loop-closure cost for forming these cluster does not depend on the sequence in which they are formed. 

These characteristic unfolding events are also reflected in the pairwise sequencing of the contacts. However, on the level of individual contacts, the parallelity of unfolding events is clearly increased. Our results thus illustrate that a highly parallel picture of folding is obtained on microstructural levels \cite{Dill97}, here the level of individual contacts. An obvious reason for this increase is that an unfolding trajectory on the contact level is specified by an opening sequence of 69 contacts. On the cluster level, trajectories are specified by a sequence of only four elements, the contact clusters. Another reason is that parallel events occur also within clusters on the contact level, not only between clusters.  A pairwise correlation analysis of the unfolding times of the contacts shows a high degree of correlation within the contact clusters.  The contact clusters thus represent cooperative substructures.

\section*{Model and methods}

\subsubsection*{Molecular dynamics simulations}

\begin{figure}[b]
\hspace*{-0.2cm}
\resizebox{1.2\linewidth}{!}{\includegraphics{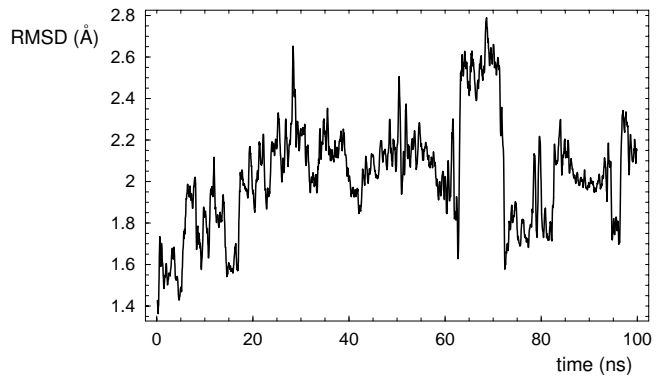}}
\caption{Root-mean-square deviation (RMSD) of the C$_\alpha$ atoms with respect to the crystal structure along a 100 nanosecond MD-trajectory at the temperature 300 K. The average value of 2.0 \AA\  for the C$_\alpha$-RMSD indicates that the protein CI2 is stable in the  CHARMM  force field EEF1 at 300 K on this timescale. The RMSD values here are averaged over time intervals of 250 ps to integrate out small-time-scale fluctuations. During unfolding at high temperatures between 400 and 500 K, the C$_\alpha$-RMSD attains values of 12 \AA\ and higher (data not shown).}
\label{figure_rmsd300K}
\end{figure}

We have performed MD simulations with the CHARMM EFF1 force field \cite{Lazaridis99,Brooks83}. EEF1 is a force field with implicit solvent \cite{Feig04} and has been previously used by several groups to study the unfolding kinetics of proteins \cite{Lazaridis97,Sham02,Ma03,Morra03}, including the protein CI2 \cite{Lazaridis97}.  After minimization of the CI2 crystal structure (protein data bank code: 2ci2) by 1000 steepest-descent and 1000 adopted-Newton-Raphson minimization steps and after 5 nanoseconds (ns) of equilibration, we have performed a 100 ns simulation at the temperature 300 K. The average root-mean-square deviation (RMSD) of the $C_\alpha$ atoms in the simulation with respect to the crystal structure was 2.0 \AA, { see Fig.~\ref{figure_rmsd300K}}. We took 50 conformations from this trajectory as starting conformations for the thermal unfolding simulations at 400 K, 450 K and 500 K. The length of the individual unfolding simulations depended on the vanishing of the native contacts and was about  100 ns at 400 K, 10 ns at 450 K, and 1 ns at 500K. We have performed 30 unfolding simulations at 400 K, and 50 unfolding simulations at 450 and 500 K. 

\subsubsection*{Unfolding times and sequences}

To define the unfolding times for a specific contact on a trajectory at 400 K, we consider time intervals of length 150 ps and determine the probability that the contact is formed during this interval. The unfolding time of this contact is defined as the time at which the probability first falls below the threshold value 0.05. In other words, the unfolding time of a contact is defined as the midpoint of the first 150 ps interval during which the contact was only present 5\% of the time. For the trajectories at 450 and 500 K, we use shorter time intervals of length 54 ps and 10.5 ps, respectively, to define contact unfolding times. We consider here as native contacts all contacts that were present during at least 75\% of an exemplary trajectory at 300 K (see Fig.~\ref{figure_contactmap}). In a given conformation, two residues were taken to be in contact if the distance between their C$_\alpha$ or C$_\beta$ atoms was less than 6~\AA. \footnote{This contact criterion also has been used in previous work\cite{Weikl03a,Weikl05}. Other widely used contact criteria involve only the C$_\alpha$ atoms, or all non-hydrogen atoms of the residues. Our criterion is relatively restrictive in the sense that the resulting number of contacts for a protein structure is relatively small. The criterion leads to 69 native contacts for CI2, i.e.~69 contacts that where present during at least 75\% of a trajectory at 300 K.  The non-hydrogen-atom criterion with a relatively small cut-off distance of 4~\AA, for example, leads to 71 contacts. According to the non-hydrogen-atom criterion, two residues are in contact if the separation of any of their non-hydrogen atoms is smaller than the cut-off distance.
}

\begin{figure}
\begin{center}
\resizebox{\linewidth}{!}{\includegraphics{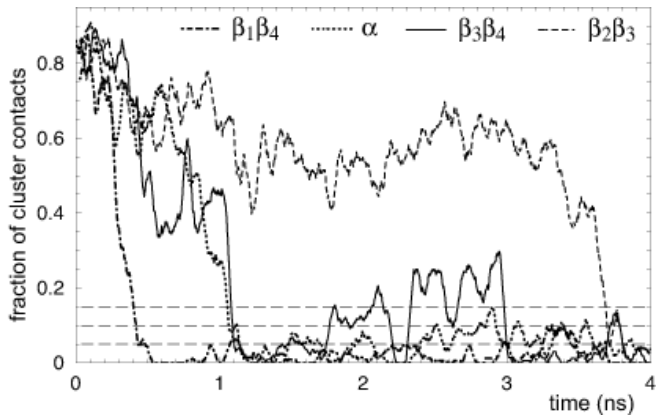}}
\end{center}
\caption{Fraction of cluster contacts on an exemplary unfolding trajectory at the temperature 450 K. The four contact clusters $\alpha$, $\beta_1\beta_4$, $\beta_2\beta_3$, and $\beta_4\beta_5$ of CI2 are defined in Fig.~\ref{figure_contactmap}. The contact fractions are averaged over time intervals of 50 ps to integrate out small-time-scale fluctuations. The three dashed horizontal lines represent thresholds used to define the unfolding sequence of the contact cluster. In this example, $\beta_1\beta_4$ is defined to unfold first because its contact fraction crosses all threshold lines prior to the contact fractions of the other clusters. By the same definition, the cluster $\beta_2\beta_3$ here unfolds last. In this example, the clusters $\alpha$ and $\beta_3\beta_4$ unfold `simultaneously' since  $\alpha$ crosses the threshold lines at the contact fraction 0.15 and 0.1 earlier than $\beta_3\beta_4$, but threshold line at 0.05 later.}
\label{figure_trajectory}
\end{figure}

Besides contacts, we consider here contact clusters as coarser structural level. The four contact clusters of the protein CI2 correspond to the $\alpha$-helix and the three strand pairings $\beta_1\beta_4$, $\beta_2\beta_3$, and $\beta_3\beta_4$ (see Fig.~\ref{figure_contactmap}). For each of the clusters, we determine the fraction of cluster contacts formed during a trajectory (see Fig.~\ref{figure_trajectory}). To define the unfolding sequence of clusters on a trajectory, we consider several threshold values for the fraction of cluster contacts. If two clusters unfold more or less simultaneously, the sequence in which they cross different threshold values can vary. We define a cluster to unfold before another cluster if it crosses all threshold values before that cluster. We have considered here 7 threshold values  between 0.05 and 0.2, in intervals of 0.025.

\begin{figure}
\begin{center}
\resizebox{0.5\linewidth}{!}{\includegraphics{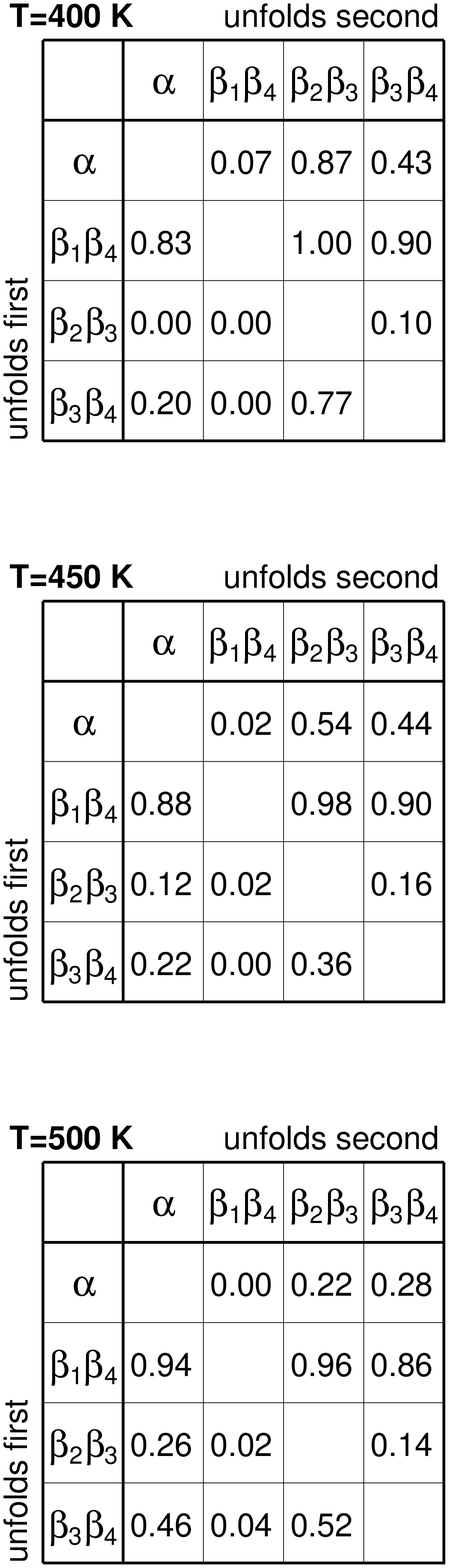}}
\end{center}
\caption{Unfolding statistics for the contact clusters of CI2 at the temperatures 400 K, 450 K, and 500 K. The numbers represent the fraction of unfolding trajectories on which contact cluster $x$ unfolds prior to contact cluster $y$ at all considered thresholds (see Fig.~\ref{figure_trajectory}). At 400 K, for example, the cluster $\beta_1\beta_4$ opens prior to $\alpha$ on 83\% of the trajectories, and  $\alpha$ opens prior to $\beta_1\beta_4$  on 7\% of the trajectories. On the remaining 10\% of trajectories, the two clusters unfold `simultaneously', i.e.~the sequence of unfolding events depends on the considered threshold.}
\label{figure_clustersequences}
\end{figure}

\section*{Results and discussion}

\subsubsection*{Parallel and sequential unfolding of contact clusters}

A statistical analysis of the unfolding events on the contact cluster level is presented in Fig.~\ref{figure_clustersequences}. The numbers indicate the fractions of trajectories on which a given cluster unfolds prior to another cluster. At the temperature 400 K, for example, $\beta_1\beta_4$ unfolds prior to $\alpha$ on 83\% of the trajectories, and $\alpha$ unfolds before  $\beta_1\beta_4$ on 7\% of the trajectories. On the remaining 10\% of trajectories, the two clusters unfold simultaneously, i.e.~without clear sequence.

\begin{figure}
\caption{Unfolding statistics for native cluster contacts of CI2 at 400 K, 450 K, and 500 K. The colors represent the fraction of unfolding trajectories on which contact $x$ unfolds prior to contact $y$. For the cluster $\alpha$, the contacts are arranged in the order 12/16, 13/16, 14/17, 15/18, 16/19, 17/20, 18/21, 18/22, 19/22, 20/23, 21/24, 21/15, 22/25 (`N to C terminus'). For $\beta_1\beta_4$, the contacts are arranged as 14/56, 13/56, 13/57, 12/57, 11/57, 12/58, 11/58, 10/58, 9/58, 10/61, 9/62, 7/61, 6/62, 5/62, 5/63, 4/64 (increasing contact order). For $\beta_2\beta_3$, the order of contacts is 28/46, 28/48, 29/46, 29/47, 30/48, 31/49, 32/50, 32/52, 33/49, 33/50, 33/51, 33/52, 34/52, 34/54, 35/51, 35/52, 35/53. For $\beta_3\beta_4$, the contacts are arranged in the order 52/56, 53/57, 52/57, 52/58, 53/59, 51/58, 52/59, 50/58, 51/60, 50/62, 49/62, 49/63, 48/64, 47/64, 47/65, 45/64 (increasing contact order).}
\label{figure_contactsequences}
\end{figure}
\begin{figure}
\begin{center}
\vspace*{-1.2cm}
\resizebox{0.85\linewidth}{!}{\includegraphics{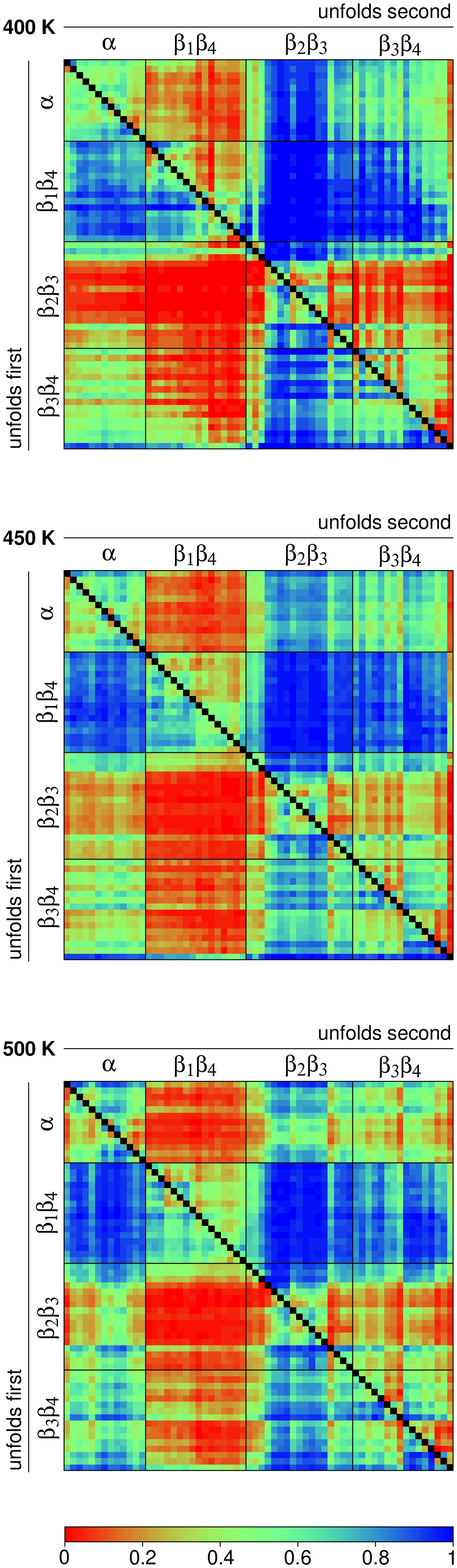}}
\end{center}
\end{figure}

At all three temperatures, the cluster $\beta_1\beta_4$ unfolds with high probability prior to the other clusters (the numbers in the second row of the matrices are between 0.83 and 1), and unfolds with low probability after the other clusters (the numbers in the second column are between 0 and 0.07). The unfolding of $\beta_1\beta_4$ with respect to each of the other three clusters thus is sequential. In contrast, the two clusters $\alpha$ and $\beta_3\beta_4$ unfold in parallel. At the temperatures 400 K and 450 K, $\beta_3\beta_4$ unfolds prior to $\alpha$ with probabilities around 0.2, and after $\alpha$ with probabilities slightly larger than 0.4. On the remaining close to 40\% of trajectories, the two clusters unfold simultaneously. At these temperatures, the unfolding of the two clusters is parallel with a 2 to 1 preference for $\alpha$ unfolding prior to $\beta_1\beta_4$ on the trajectories where the two clusters do not unfold simultaneously.  At the temperature 500 K, the unfolding of $\alpha$ and $\beta_1\beta_4$ is parallel with reversed preferences.

We observe a more pronounced temperature dependence for the unfolding sequences of $\beta_2\beta_3$ and $\alpha$. The cluster $\beta_2\beta_3$ unfolds sequentially after $\alpha$ at 400 K, and parallel to $\alpha$ at 500 K. Similarly, the unfolding of  $\beta_2\beta_3$ and $\beta_3\beta_4$ has a stronger parallel character at the higher temperatures 450 K and 500 K compared to 400 K, with a preference for $\beta_3\beta_4$ opening first.

The overall picture emerging from these statistics is: (1) $\beta_1\beta_4$ unfolds prior to the other three clusters, and (2) the three clusters $\alpha$, $\beta_2\beta_3$ and $\beta_3\beta_4$ unfold predominantly parallel to each other, with increasing parallelity at higher temperatures. This picture is in agreement with a simple folding model based on the loop-closure dependencies between the clusters \cite{Weikl03a,Weikl05}. In this model, the length of the loop that has to be closed to form a contact cluster is estimated via the graph-theoretical concept of effective contact order (ECO) \cite{Fiebig93,Dill93}. The ECO is the length of the shortest path between two residues that are forming a contact. The elementary steps along this path either are covalent contacts between neighboring residues in the chain, or previously formed noncovalent contacts between residues. In the case of the protein CI2, the ECO of the highly nonlocal cluster $\beta_1\beta_4$ is reduced by the previous formation of the other three clusters, whereas the ECOs of these three clusters do not depend on the formation of other clusters. On the dominant folding route with smallest loop-closure cost,  $\beta_1\beta_4$ is therefore predicted to form after the other three clusters, which fold in parallel. This folding sequence is reversed compared to the MD unfolding sequence. { However, we can not exclude that our MD unfolding sequence may also result from on average weaker contact energies for the $\beta_1\beta_4$ contact cluster, compared to the other three clusters.}

\begin{figure}[t]
\caption{Spearman correlation coefficients for the unfolding times of contacts. High correlations between pairs of contacts are represented in black, low correlations in white. High correlations are observed  predominantly between contacts of the same contact cluster. The contact clusters thus correspond to cooperative protein substructures. The contacts are presented in the same order as in Fig.\ 5. -- For the 30 trajectories at 400 K, a Spearman correlation coefficient of 0.43 has a $p$-value of 0.01, and a correlation coefficient 0.55 a $p$-value of 0.001 \cite{Zar72}. For the 50 trajectories at 450 and 500 K, the correlation coefficients 0.33 and 0.43 have the $p$-values 0.01 and 0.001, respectively. The $p$-value of a correlation coefficient is the probability that a similar or higher correlation is obtained by chance. The $p$-value is a measure for the significance of an observed correlation coefficient. Low $p$-values indicate high significance.}
\label{figure_correlation}
\end{figure}

\begin{figure}
\begin{center}
\resizebox{0.86\linewidth}{!}{\includegraphics{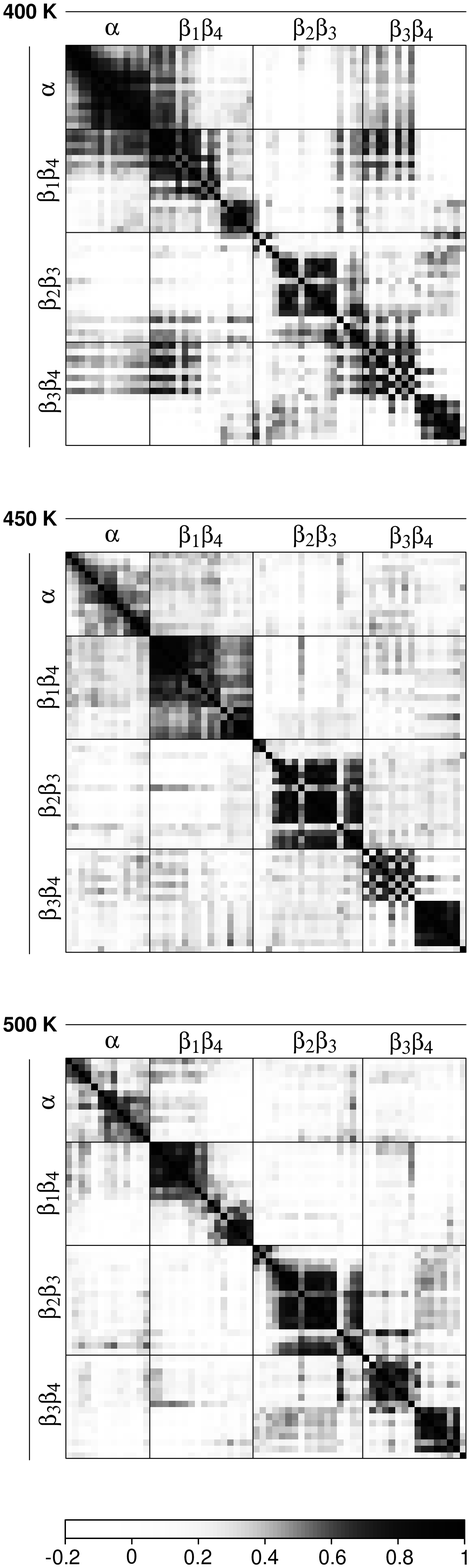}}
\end{center}
\end{figure}

\subsubsection*{Unfolding sequences and correlations of individual contacts}

The unfolding statistics of all pairs of cluster contacts is summarized in Fig.~\ref{figure_contactsequences}. The precise order in which the contacts of the four clusters are presented is specified in the figure caption. Blue colors indicate high probabilities for unfolding sequencies, and red colors low probabilities. Green colors represent intermediate probabilities, which correspond to parallel events. At all three temperatures, the contacts of $\beta_1\beta_4$ unfold with high probabilities prior to the contacts of the other clusters. At the temperature 400 K, the majority of $\beta_2\beta_3$ contacts have a strong tendency to unfold after the contacts of the clusters $\alpha$ and $\beta_3\beta_4$. This tendency decreases with increasing temperature. The unfolding statistics on the level of individual contacts thus reflects the parallel and sequential events on the cluster level. 

The correlations of the contact unfolding times are represented in Fig.~\ref{figure_correlation}. The contacts are given in the same order as in Fig.~\ref{figure_contactsequences}. Here, black indicates high correlations, and white low correlations. The correlations are quantified by the Spearman rank correlation coefficient. To calculate the Spearman coefficient, one has to consider the pairs of unfolding times $(a_1,b_1)$,  $(a_2,b_2)$, $\ldots$,  $(a_N,b_N)$ of two contacts A and B from all $N$ trajectories. The unfolding times $a_i$ of contact $A$ are then ranked according to their magnitude, and the unfolding times $b_i$ of contact $B$ as well. The Spearman rank correlation coefficient is defined as
\begin{equation}
r = 1 - 6\sum_{i=1}^N \frac{d_i^2}{N(N^2-1)}
\end{equation}
where $d_i$ is the rank difference between $a_i$ and $b_i$. The Spearman rank correlation can attain values between $-1$ and 1, with 1 representing perfect correlation, and $-1$ perfect anticorrelation. A value of 1 is obtained if the smallest unfolding time of contact $A$ is paired with the smallest unfolding time of contact $B$, the next-smallest unfolding time of $A$ with the next-smallest unfolding time of $B$, etc. The rank difference $d_i$ of all pairs of unfolding times then is zero. The Spearman correlation coefficient is a simple analogue of the Pearson correlation coefficient. The Spearman correlation is preferable here because it is less sensitive to outliers and, hence, is a more robust measure of correlation.

We obtain high correlations mostly between contacts of the same  contact cluster. The correlations of these contacts are represented in the four sub-matrizes along the diagonal of the matrizes. The high correlations indicate a high degree of substructural cooperativity within contact clusters. For the cluster $\alpha$, these correlations decrease with increasing temperature. For the clusters $\beta_1\beta_4$ and $\beta_3\beta_4$, the high correlations between contacts of the same cluster mostly appear in two `sub-blocks'. The contacts of these clusters are ordered according to increasing contact order (see caption of Fig.~\ref{figure_contactsequences}). The contact order of a contact between residues $i$ and $j$ simply is the sequence separation $|i-j|$. In the contact map shown in Fig.~\ref{figure_contactmap}, the cluster $\beta_1\beta_4$ has a small gap between the contacts 14/56 to 9/58 with smaller contact order and the contacts 10/61 to 4/64 with slightly larger contact order. A similar gap also appears in cluster $\beta_3\beta_4$. The two sub-blocks in the correlations  between $\beta_1\beta_4$ contacts correspond to high correlations within each of the two groups of contacts. This is also the case for the two sub-blocks in the $\beta_3\beta_4$ correlations. A comparison with Fig.~\ref{figure_contactsequences} also reveals a tendency for `zipping' in $\beta_1\beta_4$ and $\beta_3\beta_4$, i.e.~contacts with higher contact order in these clusters have a tendency to unfold earlier than contacts with lower contact order. This can be seen from the dominance of blue colors below the diagonals and red colors above the diagonals of the sub-matrizes in Fig.~\ref{figure_contactsequences} that represent the unfolding statistics within the $\beta_1\beta_4$ and $\beta_3\beta_4$ cluster. 

\subsubsection*{Comparison with MD simulation results of other groups}

A statistical analysis of MD unfolding sequences of the protein CI2 has also been performed by Lazaridis and Karplus \cite{Lazaridis97} and Ferrara et al.\  \cite{Ferrara00a,Ferrara00b}. Lazaridis and Karplus \cite{Lazaridis97} have considered the average times for the last appearance of contacts in unfolding simulations of CI2 at the temperature 500 K. They found the smallest average times for contacts between $\beta_1$ and $\beta_4$, the next-largest average times for contacts between $\beta_3$ and $\beta_4$ and for contacts within the $\alpha$-helix, and obtained the largest average times for contacts between $\beta_2$ and $\beta_3$. Ferrara et al.\ \cite{Ferrara00a,Ferrara00b} have considered the average C$_\alpha$ RMSDs of conformations for which groups of contacts disappeared first and appeared last. The C$_\alpha$ RMSD with respect to the native state here served as progress variable for unfolding. Ferrara et al.\ found the smallest average RMSD values at disappearance, i.e.~early unfolding, for the $\beta_1\beta_4$ and $\beta_3\beta_4$ contact groups, followed by RMSD values for the $\beta_2\beta_3$ contact groups, and obtained the largest average RMSD values at disappearance of the contacts of the $\alpha$-helix. The on average early unfolding of $\beta_1\beta_4$ observed by the two groups is in agreement with our results.

However, our analysis can not be directly compared to sequences of average unfolding times. We identify on each trajectory the unfolding sequences of pairs of contacts or contact clusters, and subsequently estimate probabilities  for particular sequences from the numbers of times these sequences appear among all trajectories. The purpose of this analysis is to determine characteristic parallel and sequential unfolding events. Average unfolding times do not reveal this information. For example, a larger average unfolding time for contact A than for contact B is observed if this contact unfolds after contact B on all trajectories (sequential unfolding), but can also be obtained if contact A opens after contact B on some trajectories, and prior to contact B on other trajectories (parallel unfolding).

In our analysis, we have focused on parallel and sequential events, and have not considered transition states for unfolding. The reason is that the unfolding scenario at the high temperatures considered here is not a  two-state scenario, but rather resembles a `downhill-unfolding' scenario. In such a scenario, the initial state of the simulation, the folded state, is instable rather than metastable, see  Fig.~\ref{figure_trajectory}.  The unfolding process is then downhill in free energy and does not involve the crossing of a significant transition-state barrier. 
Putative transition state structures have been extracted from high-temperature simulations with a conformational clustering method \cite{Li96,Day05}.  At lower temperatures, two-state folding and unfolding has been observed in MD simulations of peptides and  small mini-proteins \cite{Ferrara00,Zagrovic02,Snow02,Snow04,Settanni05,Cavalli05}.

\section*{Conclusions}

We have quantified  the degree of sequentiality for pairs of contacts and contact clusters during thermal unfolding of CI2. On the level of contact clusters, the characteristic sequential event is the unfolding of  $\beta_1\beta_4$ prior to the clusters $\alpha$, $\beta_2\beta_3$, and $\beta_3\beta_4$. The unfolding of these other three clusters is predominantly parallel. This unfolding scenario is also reflected on the contact level. The structural level of individual contacts is comparable to the microstate level of simpler statistical-mechanical models for protein folding. On this level, the unfolding process is highly parallel because of the large number of viable unfolding sequences of the 69 contacts.  A correlation analysis of the unfolding times of the contacts reveals high correlations predominantly within contact clusters. The contact clusters thus correspond to cooperative protein substructures. Experimentally, cooperative substructures have been recently observed during cold unfolding of the protein Ubiquitin \cite{Babu04} and in equilibrium and kinetic hydrogen exchange studies of Cytochrome C \cite{Maity05}.

\end{document}